\documentclass[letterpaper, 10pt, conference]{IEEEtran}
\IEEEoverridecommandlockouts
\usepackage[letterpaper, top=60pt, bottom=55pt, left=48pt, right=48pt]{geometry}
\usepackage{cite}
\usepackage{amsmath,amssymb,amsfonts}
\usepackage{algorithmic}
\usepackage{graphicx}
\usepackage{textcomp}
\usepackage{xcolor}
\usepackage{float}
\usepackage{booktabs}
\usepackage{subfigure}
\usepackage{amsmath}
\usepackage{epstopdf}
\usepackage{epsfig}
\usepackage[ruled]{algorithm2e}
\usepackage{multirow}
\usepackage{hyperref}
\usepackage{xcolor}
\usepackage{colortbl} 
\usepackage{hhline} 
\usepackage{longtable}
\usepackage{lscape}
\usepackage{float}
\hypersetup{hidelinks}

\DeclareMathOperator{\Log}{Log}
\def\BibTeX{{\rm B\kern-.05em{\sc i\kern-.025em b}\kern-.08em
		T\kern-.1667em\lower.7ex\hbox{E}\kern-.125emX}}
\begin{document}

\title{Uncertainty-Aware 3D UAV Tracking Using Single-Anchor UWB Measurements}

\title{Uncertainty--Aware Single--Anchor UWB 3--D UAV Tracking}

\author{Yuqi Ping, Junwei Wu, Bofeng Zheng, Fan Liu, Tianhao Liang and Tingting Zhang
	\thanks{Yuqi Ping, Junwei Wu, Bofeng Zheng, Fan Liu, Tianhao Liang and Tingting Zhang are with the College of Informatics, Harbin Institute of Technology, Shenzhen 518000, China (e--mail: pingyq@stu.hit.edu.cn, 220210419@stu.hit.edu.cn, zbf031018@163.com, liufan0613@stu.hit.edu.cn, liangth@hit.edu.cn, zhangtt@hit.edu.cn).
	}
}
\maketitle

\begin{abstract}
	In this letter, we present an uncertainty-aware single-anchor Ultra-Wideband (UWB)-based 3D tracking framework. Specifically, a mobile Unmanned Aerial Vehicle (UAV) maintains a desired standoff distance to a moving target using range and 3D bearing measurements from a multi-antenna UWB anchor rigidly mounted on the UAV. To enhance the stability and safety under measurement degradation and motion uncertainty, we jointly design a robust factor-graph-based target localization method and a covariance-aware control Lyapunov function--control barrier function (CLF--CBF) tracking controller. This controller adaptively adjusts distance bounds and safety margins based on the posterior target covariance provided by the factor graph. The proposed system is evaluated through numerical simulations and real-world experiments carried out in a narrow indoor corridor environment.
\end{abstract}
\section{Introduction}
Indoor deployments of Unmanned Aerial Vehicles (UAVs) and mobile robots are rapidly expanding in logistics \cite{motroni2020sensor}, manufacturing \cite{leong2024exploring}, inspection \cite{jiang2021fault}, and emergency response \cite{niroui2019deep}, making autonomous target-following increasingly routine yet safety-critical \cite{eirale2025human}. In many infrastructure-free and Global Navigation Satellite System (GNSS)-denied indoor spaces, tracking must be carried out in cluttered, space-constrained environments with frequent occlusions while maintaining a safe standoff distance. However, vision- and LiDAR-based relative localization often degrades under Non-Line-of-Sight (NLoS) conditions, poor lighting, or airborne obscurants such as dust or smoke, leading to unstable closed-loop control \cite{zhao2024subt,hu2023tightly}. These challenges motivate sensing and control frameworks capable of reliable and safe tracking under severe perception degradation.

Ultra-Wideband (UWB) provides nanosecond-level timing resolution and strong multipath resilience, enabling accurate ranging in complex indoor environments \cite{11142811}. Recent advances demonstrate its potential for UAV and mobile robot localization in GNSS-denied scenarios \cite{10637959}. Representative UWB-based tracking approaches include multi-anchor localization with velocity-based controllers \cite{feng2018human,bae2022component,santoro2022catch}, observer-based human path-following \cite{deremetz2020autonomous}, and fusion with vision or LiDAR \cite{sarmento2024fusion,zhai2024mobile,janousek2024target}. However, most existing solutions require multiple anchors or additional exteroceptive sensors, limiting their fast deployment on compact platforms.

Motivated by the demand for low-cost infrastructures, single-anchor UWB localization based on antenna arrays has emerged as a promising solution. By leveraging multiple antennas, a single anchor can estimate relative range and a 3D bearing vector via Time of Arrival (ToA) and Angle of Arrival (AoA) measurements. Prior studies have sought to enhance measurement accuracy during the signal processing stage by employing methods such as hybrid ToA and Phase Difference of Arrival (PDoA) schemes \cite{9440892} or machine-learning-based Channel Impulse Response (CIR) processing \cite{zeng2024robust,zhao2021learning}. Nevertheless, inherent constraints such as limited array aperture, antenna coupling, and multipath propagation often induce heavy-tailed, time-varying AoA errors. Compounded by a lack of geometric redundancy that hinders effective outlier rejection \cite{cao2020accurate,zhao2022finding}, single-anchor bearing estimates are prone to intermittent large deviations, thereby compromising the reliability of relative pose estimation.

From a control perspective, most UWB-based target tracking controllers treat UWB localization as a deterministic measurement and disregard uncertainty in the control design \cite{feng2018human,bae2022component,deremetz2020autonomous,zhai2024mobile,janousek2024target,kwon2024uwb}. Consequently, safety is typically enforced using heuristic margins. However, sensing uncertainty, particularly in single-anchor UWB systems, is seldom propagated to the controller. Furthermore, existing single-anchor localization approaches emphasize measurement accuracy but rarely integrate estimation and control, leaving localization uncertainty unaccounted for in tracking and safety constraints.

In this letter, we explicitly leverage perception-side uncertainty in single-anchor UWB localization for safe standoff tracking in narrow indoor environments. The main contributions are summarized as follows:
\begin{itemize}
	\item First, we propose an uncertainty-aware single-anchor UWB framework that tightly couples state estimation with safety-critical control for reliable 3D tracking.
	\item Second, we develop a robust factor graph estimator fusing UWB range and 3D bearing measurements, and design a covariance-aware controller based on Control Lyapunov Functions (CLFs) and Control Barrier Functions (CBFs) that incorporates estimation uncertainty through confidence-tube deadzones and tightened safety constraints.
	\item Finally, we validate the system through numerical simulations to demonstrate its robustness against measurement outliers and safety compliance, and deploy the complete pipeline on a UAV for real-world experiments in narrow indoor corridors with dynamic occlusions.
\end{itemize}

\section{System Model and Problem description}

\begin{figure}[h]
	\centering
	\includegraphics[trim=0 10 16 19, clip,width=0.45\textwidth]{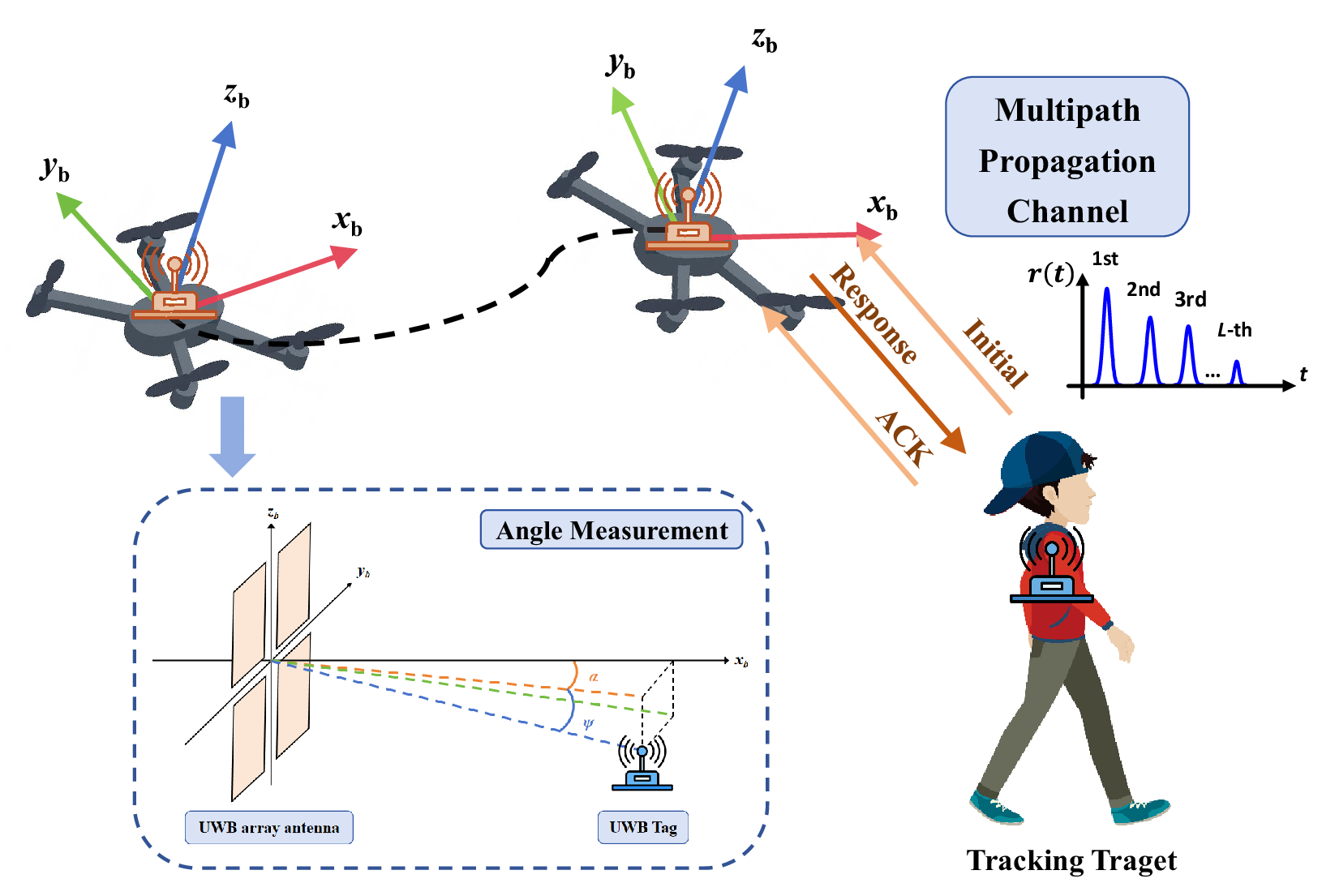}
	\caption{Illustration of the considered tracking scenario.}
	\label{fig:scenario_diagram}
\end{figure}

We consider a 3D tracking scenario, illustrated in Fig.~\ref{fig:scenario_diagram}, in which a UAV equipped with a multi-antenna UWB anchor follows a moving target carrying a UWB tag. The anchor is mounted on the UAV and the tag is attached to the target. We assume the anchor frame, referred to as the sensor frame $S$, coincides with the UAV body frame, and the tag frame coincides with the target frame. Let $E$ denote the global frame. The UAV pose at time step $k$ is assumed known and is represented by its position $\mathbf{p}_{\mathrm{R},k}\in\mathbb{R}^3$ and orientation $\mathbf{R}_{SE,k}\in SO(3)$, which transforms vectors from $E$ to $S$.

Leveraging the antenna array, the UAV performs simultaneous ranging and angle measurement with the target over a multipath propagation channel. In this section, we first introduce the dynamics models of the UAV and the target, together with the UWB measurement models, and then define the estimation and control problem considered.

\subsection{UAV Dynamics Model}

The UAV is modeled as a discrete-time double integrator
\begin{equation}
	\mathbf{X}_{\mathrm{R},k+1} = \mathbf{A}\mathbf{X}_{\mathrm{R},k} + \mathbf{B}\mathbf{u}_{\mathrm{R},k},
\end{equation}
where $\mathbf{X}_{\mathrm{R},k} = [\mathbf{p}^\top_{\mathrm{R},k},\mathbf{v}^\top_{\mathrm{R},k}]^\top \in \mathbb{R}^6$
denotes the UAV position and velocity. Given the sampling time $\Delta t$, the state transition and control matrices are defined as
\begin{equation}
	\mathbf{A} =
	\begin{bmatrix}
		\mathbf{I} & \Delta t\,\mathbf{I} \\
		\mathbf{0} & \mathbf{I}
	\end{bmatrix},\quad
	\mathbf{B} =
	\begin{bmatrix}
		\tfrac{1}{2}\Delta t^2\,\mathbf{I} \\
		\Delta t\,\mathbf{I}
	\end{bmatrix}.
\end{equation}
The acceleration input $\mathbf{u}_{\mathrm{R},k}\in\mathbb{R}^3$ is subject to actuator saturation and velocity constraints
\begin{equation}
	\mathbf{u}_{\min} \le \mathbf{u}_{\mathrm{R},k} \le \mathbf{u}_{\max}, \quad
	\|\mathbf{v}_{\mathrm{R},k}\|_\infty \le v_{\max}.
\end{equation}

\subsection{Target Dynamics Model}

The target motion is modeled as a double integrator subject to unknown inputs
\begin{equation}
	\mathbf{X}_{\mathrm{T},k+1} = \mathbf{A}\mathbf{X}_{\mathrm{T},k} + \mathbf{n}_{\mathrm{T},k},
\end{equation}
where the state $\mathbf{X}_{\mathrm{T},k} = [\mathbf{p}^\top_{\mathrm{T},k}, \mathbf{v}^\top_{\mathrm{T},k}]^\top \in \mathbb{R}^6$ is defined similar to the UAV. The term $\mathbf{n}_{\mathrm{T},k} \sim \mathcal{N}(\mathbf{0}, \mathbf{\Sigma}_{\mathrm{T}})$ represents the process noise due to the target's unknown maneuvering acceleration. In addition, the magnitude of the target's acceleration $\mathbf{a}_{\mathrm{T},k}$ is assumed bounded by
\begin{equation}
	\lVert \mathbf{a}_{\mathrm{T},k} \rVert_2 \le a_{\max}.
\end{equation}

\subsection{UWB Ranging Measurement Model}
At time step $k$, the UWB anchor provides a scalar range measurement between the UAV and target.
Define the relative position in the global frame $E$ as
\begin{equation}
	\mathbf r_{E,k} \triangleq \mathbf p_{\mathrm T,k}-\mathbf p_{\mathrm R,k},
\end{equation}
and the corresponding distance as
\begin{equation}
	d_k \triangleq \|\mathbf r_{E,k}\|.
\end{equation}
The ranging measurement function is defined as
\begin{equation}
	h_r(\mathbf X_{\mathrm R,k},\mathbf X_{\mathrm T,k}) \triangleq d_k .
\end{equation}
The measurement model is given by
\begin{equation}
	z_{r,k}	= d_k + n_{r,k},
	\label{eq:range}
\end{equation}
where $n_{r,k}\sim\mathcal{N}(0,\Sigma_r)$ is zero-mean Gaussian noise with variance $\Sigma_r$.

\begin{figure*}[t]
	\centering
	\includegraphics[trim=0 0 0 0, clip,width=1\textwidth]{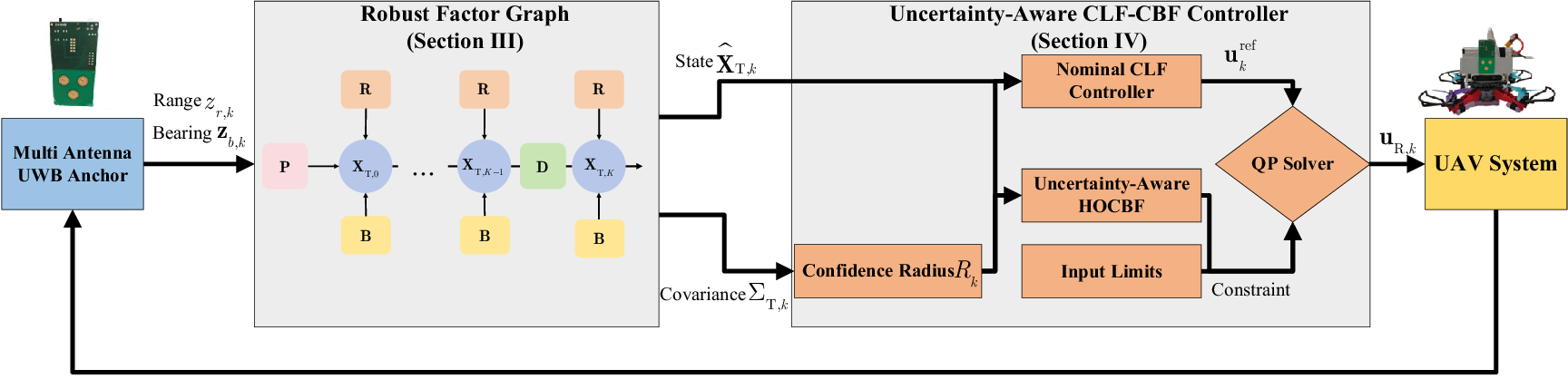}
	\caption{System architecture illustrating the integration of robust target localization and covariance-aware tracking control.}
	\label{fig:system_architecture}
\end{figure*}

\subsection{UWB 3D Angle Measurement Model}
The multi-antenna UWB anchor provides azimuth and elevation measurements of the target in the sensor frame $S$.
Using the rotation matrix $\mathbf R_{SE,k}$ and the relative position $\mathbf r_{E,k}$ defined previously, the position vector expressed in the sensor frame is
\begin{equation}
	\mathbf r_{S,k} \triangleq \mathbf R_{SE,k}\mathbf r_{E,k}
	=
	\begin{bmatrix}
		x_{S,k},y_{S,k},z_{S,k}
	\end{bmatrix}^\top .
\end{equation}

The noise-free azimuth and elevation are obtained from the spherical projection of $\mathbf r_{S,k}$ and are modeled by the measurement function
\begin{equation}
	\mathbf h_b(\mathbf X_{\mathrm R,k},\mathbf X_{\mathrm T,k}) =
	\begin{bmatrix}
		\mathrm{atan2}(y_{S,k},x_{S,k}) \\[4pt]
		\mathrm{atan2}\!\bigl(z_{S,k},\sqrt{x_{S,k}^2+y_{S,k}^2}\bigr)
	\end{bmatrix}.
\end{equation}
The corresponding noisy angle measurement is
\begin{equation}
	\mathbf z_{b,k} =
	\begin{bmatrix}
		\alpha_k \\ \beta_k
	\end{bmatrix}
	=
	\mathbf h_b(\mathbf X_{\mathrm R,k},\mathbf X_{\mathrm T,k})
	+ \mathbf n_{b,k},
\end{equation}
where, under nominal conditions, we assume Gaussian measurement noise 
$\mathbf n_{b,k}\sim\mathcal{N}(\mathbf 0,\boldsymbol{\Sigma}_b)$.
In practical UWB deployments, multipath propagation and NLoS effects may occasionally induce large angular deviations that violate the Gaussian assumption, resulting in outliers during angle measurements.

\subsection{Problem Description}

Based on the dynamics and sensing models formulated above, the UAV receives noisy UWB range and 3D angle measurements at each time step $k$. Since the target state $\mathbf{X}_{\mathrm{T},k}$ is not directly observable, it must be estimated online by fusing the available measurements $z_{r,k}$ and $\mathbf{z}_{b,k}$ under the assumed motion model.

We formulate the tracking control objective in terms of the relative distance $d_k$ and the vertical separation. Let $z_{\mathrm R,k}$ and $z_{\mathrm T,k}$ denote the vertical components of the position vectors $\mathbf{p}_{\mathrm R,k}$ and $\mathbf{p}_{\mathrm T,k}$, respectively. The altitude error is defined as
\begin{equation}
	e_{z,k} \triangleq z_{\mathrm R,k} - z_{\mathrm T,k}.
\end{equation}

The objective is to design a real-time controller that utilizes the estimated target state to regulate the relative geometry, such that
\begin{equation}
	\lim_{k\to\infty} d_k = d^\star,
	\qquad
	\lim_{k\to\infty} e_{z,k} = 0,
	\label{eq:tracking-goal}
\end{equation}
where $d^\star$ represents the desired distance. The controller must ensure these objectives while satisfying the UAV's actuator and velocity constraints, and explicitly accounting for the uncertainty in the target state estimate. The overall system architecture is illustrated in Fig.~\ref{fig:system_architecture}.

\section{Robust Factor Graph for Target Localization}
\begin{figure}[h]
	\centering
	\includegraphics[trim=0 0 0 0, clip,width=0.50\textwidth]{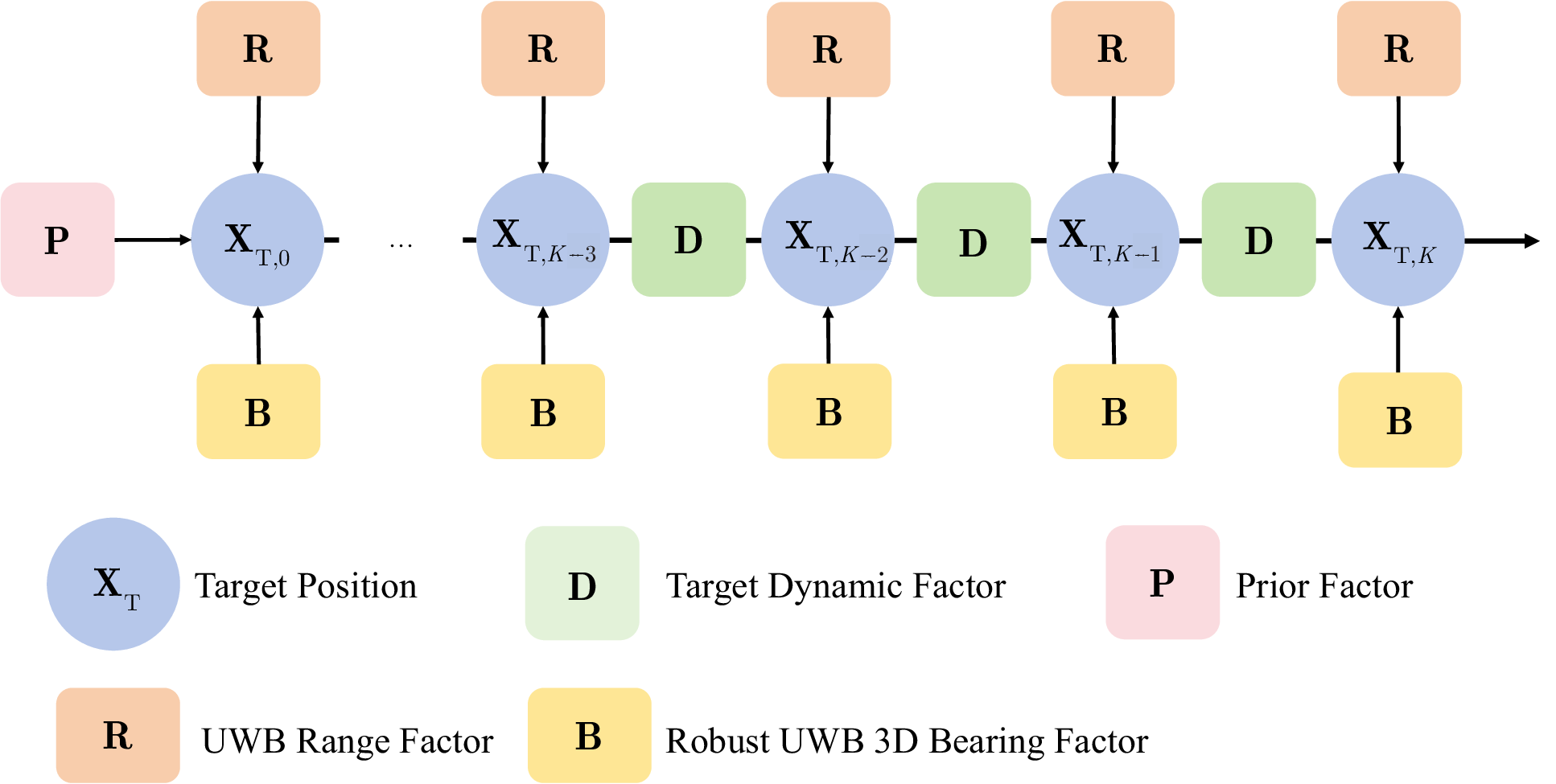}
	\caption{Structure of the proposed robust factor graph.}
	\label{fig:factor_graph}
\end{figure}

In this section, we formulate target localization as a robust factor-graph-based Maximum A Posteriori (MAP) estimation problem over the sequence of target states $\{\mathbf{X}_{\mathrm{T},k}\}_{k=0}^{K}$. As depicted in Fig.~\ref{fig:factor_graph}, the proposed factor graph fuses UWB range and 3D bearing measurements with a motion prior derived from the target dynamics.

\subsection{Prior Factor}
We define a prior factor on the initial target state to incorporate an initial estimate. The corresponding residual is
\begin{equation}
	\mathbf{r}_0
	\;=\;
	\mathbf{X}_{\mathrm{T},0}-\widehat{\mathbf{X}}_{\mathrm{T},0}.
\end{equation}
The prior factor is then expressed as a Gaussian distribution
\begin{equation}
	\phi_{\mathrm{prior}}(\mathbf{X}_{\mathrm{T},0})
	\;\propto\;
	\exp\!\left\{-\tfrac{1}{2}\,\|\mathbf{r}_0\|^2_{\boldsymbol{\Sigma}_0^{-1}}\right\}.
\end{equation}
This factor encodes the a priori belief regarding the initial state $\mathbf{X}_{\mathrm{T},0}$. The covariance matrix $\boldsymbol{\Sigma}_0$ characterizes the uncertainty of this initialization, where a larger $\boldsymbol{\Sigma}_0$ signifies lower confidence in the initial guess $\widehat{\mathbf{X}}_{\mathrm{T},0}$.

\subsection{Target Dynamic Factor}
Given the target motion model, we introduce a dynamic factor that connects consecutive target states. The residual is defined as
\begin{equation}
	\mathbf{r}^{\mathrm{dyn}}_{k}
	\;\triangleq\;
	\mathbf{X}_{\mathrm{T},k+1}
	-
	\mathbf{A}\,\mathbf{X}_{\mathrm{T},k}.
\end{equation}
The resulting dynamic factor takes the Gaussian form
\begin{equation}
	\phi_{\mathrm{dyn}}(\mathbf{X}_{\mathrm{T},k},\mathbf{X}_{\mathrm{T},k+1})
	\;\propto\;
	\exp\!\Big\{-\tfrac{1}{2}\,
	\|\mathbf{r}^{\mathrm{dyn}}_{k}\|^{2}_{\boldsymbol{\Sigma}_{{\mathrm{T}}}^{-1}}
	\Big\}.
\end{equation}

\subsection{UWB Ranging Factor}
For each valid UWB range measurement, we define a scalar residual using the range model in~\eqref{eq:range}:
\begin{equation}
	r^{\mathrm{rng}}_{k}
	\;\triangleq\;
	z_{r,k}-h_{r,k}.
\end{equation}
The corresponding Gaussian ranging factor is given by
\begin{equation}
	\phi_{\mathrm{rng}}(\mathbf{p}_{\mathrm{T},k},\mathbf{p}_{\mathrm{R},k})
	\;\propto\;
	\exp\!\Big\{-\tfrac{1}{2}\,
	\|r^{\mathrm{rng}}_{k}\|^{2}_{\Sigma_r^{-1}}
	\Big\}.
\end{equation}

\subsection{Robust UWB 3D Bearing Factor}
The UWB anchor provides bearing measurements in the form of azimuth and elevation angles. To enable a smooth residual definition on the manifold $\mathbb{S}^2$, these angular measurements are reparameterized as unit direction vectors on the sphere. The uncertainty of the angular measurements is used as an approximation of the uncertainty in the resulting 3D bearing representation. This approximation is accurate for small angular deviations and is particularly valid when the UAV and the target operate at similar altitudes, where elevation contributes minimally to bearing distortion.

From the angular measurement $\mathbf{z}_{b,k} = [\alpha_k,\beta_k]^\top$, we construct the measured 3D bearing on $\mathbb{S}^2$ as
\begin{equation}
	\mathbf{z}^{\mathbb{S}^2}_{b,k}
	=
	\begin{bmatrix}
		\cos\beta_k\,\cos\alpha_k \\
		\cos\beta_k\,\sin\alpha_k \\
		\sin\beta_k
	\end{bmatrix}
	\in \mathbb{S}^2.
\end{equation}

Given the current state estimates, the predicted (noise-free) bearing is computed geometrically as
\begin{equation}
	\mathbf{h}^{\mathbb{S}^2}_{b,k}
	=
	\dfrac{\mathbf{R}_{SE,k}
		\left(\mathbf{p}_{\mathrm{T},k}-\mathbf{p}_{\mathrm{R},k}\right)}
	{\left\|\mathbf{p}_{\mathrm{T},k}-\mathbf{p}_{\mathrm{R},k}\right\|}
	\in \mathbb{S}^2.
\end{equation}

Since both vectors lie on the unit sphere, the bearing residual is defined in the tangent space. Specifically, we project the geodesic error onto a locally defined orthonormal basis $\mathbf{B}_k \in \mathbb{R}^{3\times2}$ spanning the tangent plane:
\begin{equation}
	\mathbf{r}^{\mathrm{brg}}_{k}
	=
	\mathbf{B}_k^\top
	\Log_{\mathbf{h}^{\mathbb{S}^2}_{b,k}}\!\left(\mathbf{z}^{\mathbb{S}^2}_{b,k}\right)
	\approx
	\mathbf{B}_k^\top
	\left(\mathbf{z}^{\mathbb{S}^2}_{b,k}
	-
	\mathbf{h}^{\mathbb{S}^2}_{b,k}\right),
\end{equation}
where the approximation holds for small angular deviations.

To improve the robustness against outliers caused by multipath and NLoS effects, the residual is evaluated using a Cauchy robust loss
\begin{equation}
	\rho_{\mathrm{Cauchy}}\!\Big(
	\|\mathbf{r}^{\mathrm{brg}}_{k}\|^{2}_{\boldsymbol{\Sigma}_b^{-1}},
	c_{\mathrm{brg}}
	\Big)
	=
	c_{\mathrm{brg}}^{2}
	\log\!\left(
	1+
	\|\mathbf{r}^{\mathrm{brg}}_{k}\|^{2}_{\boldsymbol{\Sigma}_b^{-1}}
	/ c_{\mathrm{brg}}^{2}
	\right),
\end{equation}
where $c_{\mathrm{brg}}>0$ controls sensitivity to large deviations.

The resulting robust 3D bearing factor is given by
\begin{equation}
	\phi_{\mathrm{brg}}(\mathbf{p}_{\mathrm{T},k})
	\propto
	\exp\!\left(
	-\tfrac{1}{2}\,
	\rho_{\mathrm{Cauchy}}\!\Big(
	\|\mathbf{r}^{\mathrm{brg}}_{k}\|^{2}_{\boldsymbol{\Sigma}_b^{-1}},
	c_{\mathrm{brg}}
	\Big)
	\right).
\end{equation}

\subsection{Real-Time Incremental Solution}

We implement target estimation as an incremental factor graph MAP problem in GTSAM~\cite{gtsam}. At each time step $k$, we add a new target node $\mathbf{X}_{\mathrm{T},k}$ together with the corresponding prior, dynamics, UWB ranging, and robust UWB bearing factors. Let $\mathcal{X}_K\!\triangleq\!\{\mathbf{X}_{\mathrm{T},0},\ldots,\mathbf{X}_{\mathrm{T},K}\}$ denote the set of target states up to time $K$. The MAP optimization is
\begin{align}
	\mathcal{X}_K^\star
	&= \arg\min_{\mathcal{X}_K}\Big[
	\|\mathbf{r}_0\|^2_{\boldsymbol{\Sigma}_0^{-1}}
	+\sum_{k=0}^{K-1}\|\mathbf{r}^{\mathrm{dyn}}_{k}\|^{2}_{\boldsymbol{\Sigma}_{{\mathrm{T}}}^{-1}}
	+\sum_{k=0}^{K}\|r^{\mathrm{rng}}_{k}\|^{2}_{\Sigma_r^{-1}} \nonumber\\
	&\qquad\qquad
	+\sum_{k=0}^{K}\rho_{\mathrm{Cauchy}}\!\big(\|\mathbf{r}^{\mathrm{brg}}_{k}\|^{2}_{\boldsymbol{\Sigma}_b^{-1}},c_{\mathrm{brg}}\big)
	\Big].
\end{align}

We solve this problem incrementally using iSAM2~\cite{kaess2011isam2}. At each update, the nonlinear factors are linearized around the current estimate and the resulting sparse normal equations are updated in a Bayes tree, yielding an approximation of the posterior over all target states. Denoting by $\mathcal{Z}_{0:K}$ the set of all measurements up to time $K$, we approximate the joint posterior as a Gaussian
\begin{align}
	p(\mathcal{X}_K \mid \mathcal{Z}_{0:K})
	&\approx \mathcal{N}\!\big(\mathcal{X}_K^\star,\,\boldsymbol{\Sigma}_{\mathrm{post}}\big),
\end{align}
where $\mathcal{X}_K^\star$ is the MAP estimate and $\boldsymbol{\Sigma}_{\mathrm{post}}$ is the joint covariance extracted from the linearized Bayes tree representation.

From this joint Gaussian, we extract for each time step $k$ the state estimate and covariance of the target as
\begin{align}
	\widehat{\mathbf{X}}_{\mathrm{T},k}
	&\triangleq \mathbf{X}_{\mathrm{T},k}^\star
	=
	\begin{bmatrix}
		\widehat{\mathbf{p}}_{\mathrm{T},k}^\top &
		\widehat{\mathbf{v}}_{\mathrm{T},k}^\top
	\end{bmatrix}^\top
	\in\mathbb{R}^6,
	\nonumber\\
	\boldsymbol{\Sigma}_{\mathrm{T},k}
	&\triangleq
	\mathrm{Marginal}\!\big(\boldsymbol{\Sigma}_{\mathrm{post}},\,\mathbf{X}_{\mathrm{T},k}\big)
	\in\mathbb{R}^{6\times 6},
\end{align}
where $\mathbf{X}_{\mathrm{T},k}^\star$ denotes the $k$-th state component of the MAP solution $\mathcal{X}_K^\star$, and
$\mathrm{Marginal}(\cdot,\mathbf{X}_{\mathrm{T},k})$ denotes the operation that extracts the sub-block of $\boldsymbol{\Sigma}_{\mathrm{post}}$
corresponding to $\mathbf{X}_{\mathrm{T},k}$.

This incremental scheme updates the estimate online as new factors arrive, producing the target state estimate and covariance in real time. By avoiding batch re-optimization and performing incremental updates with selective relinearization over only the affected portion of the Bayes tree, iSAM2 enables real-time operation even at high measurement rates. ﻿

\section{Covariance-Aware CLF--CBF Controller}

\begin{figure}[h]
	\centering
	\includegraphics[trim=0 0 0 0, clip,width=0.37\textwidth]{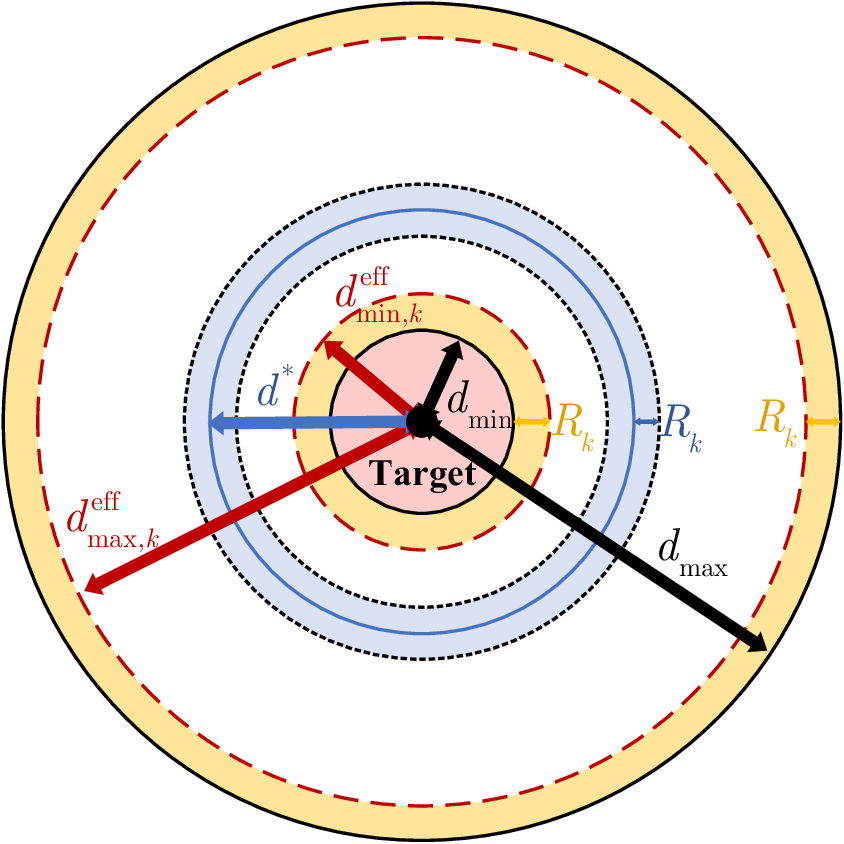}
	\caption{Covariance-aware range safety construction in which the posterior target covariance induces a confidence radius $R_k$, leading to effective near and far distance limits and an uncertainty-induced deadzone around the desired standoff distance $d^\star$.}
	\label{fig:controller}
\end{figure}

In this section, we present the covariance-aware CLF–CBF controller for 3D tracking. As illustrated in Fig.~\ref{fig:controller}, we construct a covariance-aware range safety envelope by tightening the physical distance bounds using an uncertainty radius derived from the posterior target covariance. This yields effective near and far range limits, along with an uncertainty-induced deadzone around the desired standoff distance. The detailed definitions and formulations will be provided subsequently.

\vspace{-10pt}
\subsection{Covariance-Based Uncertainty Quantification}
At each time step $k$, the robust factor graph yields a Gaussian approximation of the target posterior, summarized by the state estimate $\widehat{\mathbf{X}}_{\mathrm{T},k}$ and its covariance $\boldsymbol{\Sigma}_{\mathrm{T},k}$. To quantify the uncertainty in the estimated target position, we denote by $\boldsymbol{\Sigma}_{\mathbf{p}_{\mathrm{T}},k}\in\mathbb{R}^{3\times3}$ the position covariance, defined as the position block of $\boldsymbol{\Sigma}_{\mathrm{T},k}$. Accordingly, we construct a conservative scalar uncertainty bound via the confidence radius
\begin{equation}
	R_k \;\triangleq\; \sqrt{\chi^2_{3,\,1-\alpha}\,\lambda_{\max}\!\big(\boldsymbol{\Sigma}_{\mathbf{p}_{\mathrm{T}},k}\big)} ,
	\label{eq:rconf}
\end{equation}
where $\lambda_{\max}(\cdot)$ denotes the largest eigenvalue of a symmetric positive semidefinite matrix, $\chi^2_{3,\,1-\alpha}$ is the $(1-\alpha)$-quantile of a chi-square distribution with three degrees of freedom, and $\alpha\in(0,1)$ is a specified risk level.

Under the Gaussian posterior approximation, the position estimation error
$\mathbf{e}_k \triangleq \mathbf{p}_{\mathrm{T},k}-\widehat{\mathbf{p}}_{\mathrm{T},k}$
belongs to the confidence ellipsoid
$\{\mathbf{e}_k\in\mathbb{R}^3:\mathbf{e}_k^\top\boldsymbol{\Sigma}_{\mathbf{p}_{\mathrm{T}},k}^{-1}\mathbf{e}_k\le \chi^2_{3,1-\alpha}\}$
with probability $1-\alpha$. Accordingly, $R_k$ provides a spherical upper bound on this ellipsoid by upper-bounding its principal-axis uncertainty. In the subsequent controller design, $R_k$ is used to tighten the effective distance limits and to construct uncertainty-induced deadzones, thereby preventing aggressive reactions to deviations that are likely explained by estimation noise rather than true tracking errors.

\subsection{Covariance-Aware CLF Tracking with Confidence-Tube Deadzones}
\label{subsec:clf-ua-detailed}

Rather than relying on the inaccessible ground truth $\mathbf{r}_{E,k}$, the controller utilizes the estimated target position $\widehat{\mathbf{p}}_{\mathrm{T},k}$ and its associated relative state variables.
The estimated relative distance $\widehat d_k$ and LoS unit vector $\widehat{\mathbf{n}}_k$ are defined as
\begin{equation}
	\widehat d_k \triangleq \left\|\widehat{\mathbf{p}}_{\mathrm{T},k}-\mathbf{p}_{\mathrm{R},k}\right\|,
	\qquad
	\widehat{\mathbf{n}}_k
	\triangleq
	\frac{\widehat{\mathbf{p}}_{\mathrm{T},k}-\mathbf{p}_{\mathrm{R},k}}
	{\widehat d_k}.
\end{equation}
The estimated relative velocity $\widehat{\mathbf{v}}^{\mathrm{rel}}_k \triangleq \widehat{\mathbf{v}}_{\mathrm{T},k}-\mathbf{v}_{\mathrm{R},k}$ is decomposed into radial and tangential components
\begin{equation}
	\widehat v_{r,k}
	\triangleq
	\widehat{\mathbf{n}}_k^\top \widehat{\mathbf{v}}^{\mathrm{rel}}_k,
	\qquad
	\widehat{\mathbf{v}}_{\tau,k}
	\triangleq
	\widehat{\mathbf{v}}^{\mathrm{rel}}_k
	-
	\widehat v_{r,k}\widehat{\mathbf{n}}_k .
\end{equation}
Similarly, the estimated altitude error and its time derivative are given by
\begin{equation}
	\widehat e_{z,k}
	\triangleq
	z_{\mathrm{R},k}-\widehat z_{\mathrm{T},k},
	\qquad
	\widehat v_{z,k}
	\triangleq
	v_{\mathrm R,z,k}-\widehat v_{\mathrm T,z,k}.
\end{equation}

To prevent aggressive reactions to estimation noise, we leverage the confidence radius $R_k$ from \eqref{eq:rconf} to construct covariance-informed dead-zone errors
\begin{align}
	\tilde e_{r,k}
	&\triangleq
	\mathrm{sat}_0\!\big(|\widehat d_k-d^\star|-R_k\big)\,
	\mathrm{sign}(\widehat d_k-d^\star),
	\label{eq:er-band-det}\\
	\tilde e_{z,k}
	&\triangleq
	\mathrm{sat}_0\!\big(|\widehat e_{z,k}|-R_k\big)\,
	\mathrm{sign}(\widehat e_{z,k}),
	\label{eq:ez-band-det}
\end{align}
where $\mathrm{sat}_0(x)\triangleq\max(x,0)$. These error terms vanish when the estimated states lie within the uncertainty tube.

We adopt a quadratic CLF
\begin{align}
	V_k =
	&\tfrac12 k_r \tilde e_{r,k}^2
	+\tfrac12 k_{vr} \widehat v_{r,k}^2
	+\tfrac12 k_z \tilde e_{z,k}^2
	+\tfrac12 k_{vz} \widehat v_{z,k}^2
	\nonumber\\
	&+\tfrac12 k_\tau \|\widehat{\mathbf{v}}_{\tau,k}\|^2,
	\label{eq:clf-det}
\end{align}
where $k_r, k_{vr}, k_z, k_{vz}, k_\tau > 0$ are positive tuning gains.
In \eqref{eq:clf-det}, the position-related terms regulate the desired standoff distance and altitude, while the velocity terms provide damping in the radial and vertical directions. The tangential velocity penalty is included to align the UAV velocity with the LoS, thereby suppressing lateral relative motion.
The corresponding nominal reference acceleration, designed to exponentially stabilize the unconstrained error dynamics, is derived as
\begin{align}
	\mathbf{u}^{\mathrm{ref}}_k
	=\,
	&\big(k_r \tilde e_{r,k} + k_{vr} \widehat v_{r,k}\big)\widehat{\mathbf{n}}_k
	\nonumber\\
	&+ k_\tau\,\widehat{\mathbf{v}}_{\tau,k}
	- \big(k_z \tilde e_{z,k} + k_{vz} \widehat v_{z,k}\big)\mathbf{e}_z ,
	\label{eq:uref-det}
\end{align}
where $\mathbf{e}_z\triangleq[0,0,1]^\top$. In the absence of actuator saturation or safety constraints, $\mathbf{u}^{\mathrm{ref}}_k$ ensures that the Lyapunov derivative satisfies $\dot V_k\le -cV_k$ for some $c>0$, thus exponentially stabilizing the nominal tracking error dynamics.

\subsection{Covariance-Aware HOCBF Range Safety Constraints}

To ensure collision avoidance and reliable UWB measuement, the distance $d_k$ must remain within $[d_{\min}, d_{\max}]$. Since $d_k$ is unobservable, we enforce this constraint on the estimate $\widehat d_k$ using the confidence radius $R_k$ to define effective safety limits
\begin{equation}
	d_{\min,k}^{\mathrm{eff}} \triangleq d_{\min} + R_k,
	\qquad
	d_{\max,k}^{\mathrm{eff}} \triangleq d_{\max} - R_k.
	\label{eq:eff_margins_r}
\end{equation}
Maintaining $\widehat d_k \in [d_{\min,k}^{\mathrm{eff}},\, d_{\max,k}^{\mathrm{eff}}]$ ensures that the true distance satisfies the physical bounds with probability at least $1-\alpha$ under the Gaussian posterior assumption.

Accordingly, we define the barrier functions for the lower and upper effective bounds as
\begin{equation}
	\widehat h^{\mathrm{near}}_k
	\triangleq
	\widehat d_k - d_{\min,k}^{\mathrm{eff}},
	\qquad
	\widehat h^{\mathrm{far}}_k
	\triangleq
	d_{\max,k}^{\mathrm{eff}} - \widehat d_k .
	\label{eq:h_near_far_r}
\end{equation}
We enforce the critically damped High-Order Control Barrier Function (HOCBF) condition~\cite{xiao2021high}:
\begin{equation}
	\ddot h + 2\omega\,\dot h + \omega^2 h \ge 0,
	\qquad \omega>0.
	\label{eq:hocbf_r}
\end{equation}

Assuming $R_k$ is quasi-static within the control horizon, the second-order dynamics of the estimated distance are given by
\begin{equation}
	\ddot{\widehat d}_k =
	\widehat{\mathbf n}_k^\top(\mathbf a_{\mathrm T,k}-\mathbf u_{\mathrm R,k})
	+ \frac{\|\widehat{\mathbf v}_{\tau,k}\|^2}{\widehat d_k}.
\end{equation}
To handle the unmeasurable target acceleration $\mathbf a_{\mathrm T,k}$, we robustly satisfy \eqref{eq:hocbf_r} for the worst-case projection $|\widehat{\mathbf n}_k^\top \mathbf a_{\mathrm T,k}| \le a_{\max}$.
Substituting the dynamics into the barrier condition yields the following affine constraints on the UAV control input $\mathbf u_{\mathrm R,k}$
\begin{align}
	\widehat{\mathbf n}_k^\top \mathbf u_{\mathrm R,k}
	&\le
	-a_{\max}
	+ \frac{\|\widehat{\mathbf v}_{\tau,k}\|^2}{\widehat d_k}
	+ 2\omega\, \widehat v_{r,k}
	+ \omega^2 \widehat h^{\mathrm{near}}_k,
	\label{eq:QP-cbf-near-correct}
	\\[4pt]
	-\widehat{\mathbf n}_k^\top \mathbf u_{\mathrm R,k}
	&\le
	-a_{\max}
	- \frac{\|\widehat{\mathbf v}_{\tau,k}\|^2}{\widehat d_k}
	- 2\omega\, \widehat v_{r,k}
	+ \omega^2 \widehat h^{\mathrm{far}}_k.
	\label{eq:QP-cbf-far-correct}
\end{align}
These constraints explicitly account for the estimation uncertainty, the relative system dynamics, and the adversarial target maneuver capabilities.

\subsection{QP-Based Control with Input and Speed Constraints}

To synthesize the CLF-based tracking performance with the HOCBF safety guarantees, we formulate the control allocation as a constrained Quadratic Program (QP). At each time step $k$, the optimal control input is computed by solving:
\begin{align}
	\min_{\mathbf u_{\mathrm R,k}}\;&
	\tfrac12\big\|\mathbf u_{\mathrm R,k}-\mathbf u_k^{\mathrm{ref}}\big\|^2
	\label{eq:qp_obj}
	\\[4pt]
	\text{s.t.}\;&
	\eqref{eq:QP-cbf-near-correct},\ \eqref{eq:QP-cbf-far-correct},
	\nonumber\\
	& \mathbf u_{\min} \le \mathbf u_{\mathrm R,k} \le \mathbf u_{\max},
	\label{eq:qp_input}
	\\
	& \big\|\mathbf v_{\mathrm R,k}+\Delta t\,\mathbf u_{\mathrm R,k}\big\|_\infty \le v_{\max}.
	\label{eq:qp_speed}
\end{align}

The quadratic objective \eqref{eq:qp_obj} acts as a safety filter by minimizing the deviation of the actual control input $\mathbf u_{\mathrm R,k}$ from the nominal reference $\mathbf u_k^{\mathrm{ref}}$. This formulation prioritizes CLF-based tracking performance while enforcing minimal necessary corrections to satisfy safety constraints. Specifically, the covariance-aware HOCBF constraints \eqref{eq:QP-cbf-near-correct} and \eqref{eq:QP-cbf-far-correct} confine the estimated distance $\widehat d_k$ within the uncertainty-tightened envelope, thereby ensuring the true distance respects physical bounds with probability at least $1-\alpha$. Furthermore, physical platform limitations are enforced by \eqref{eq:qp_input} for actuator saturation and by \eqref{eq:qp_speed} for the maximum flight speed $v_{\max}$. The resulting optimization problem is a convex QP with linear inequalities, enabling efficient real-time solution for robust and safe tracking. In our implementation, this convex QP is solved online using the Operator Splitting Quadratic Program (OSQP) solver.

\begin{figure}[h]
	\centering
	\includegraphics[trim=20 30 20 30, clip,width=0.5\textwidth]{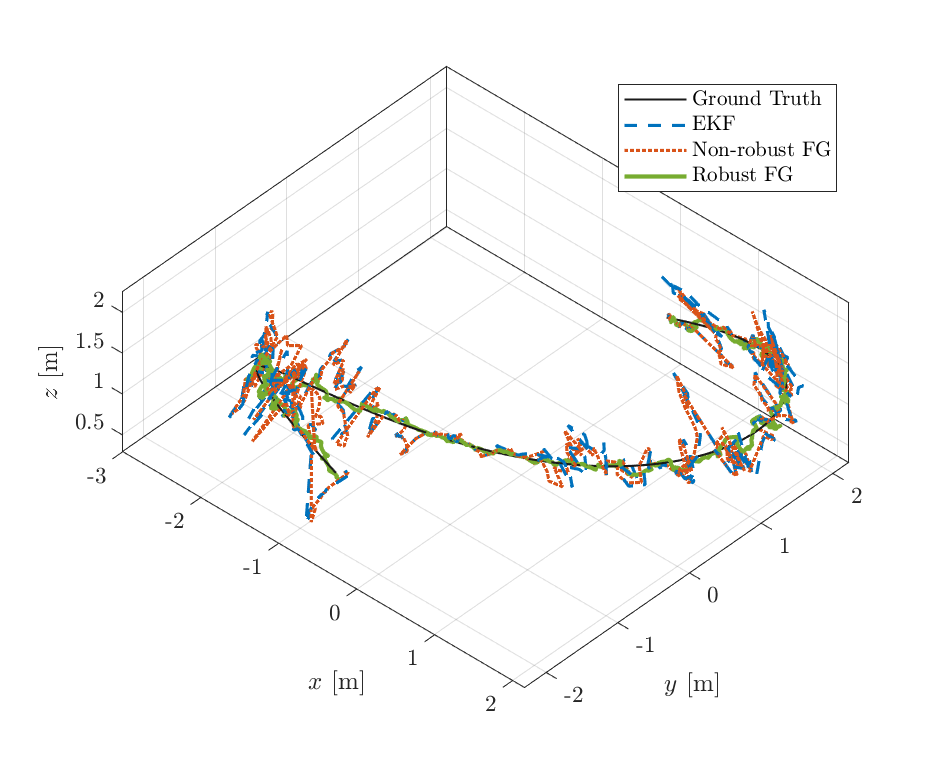}
	\caption{Representative trajectory comparison under bearing outliers ($p_\text{out}=0.2$).}
	\label{fig:sim_traj}
\end{figure}
\section{Numerical Simulation}

To validate the proposed system, we conducted numerical simulations focusing on both the robust factor graph estimator and the covariance-aware CLF--CBF controller in this section.

\begin{figure}[h]
	\centering
	\includegraphics[trim=5 0 15 10, clip,width=0.5\textwidth]{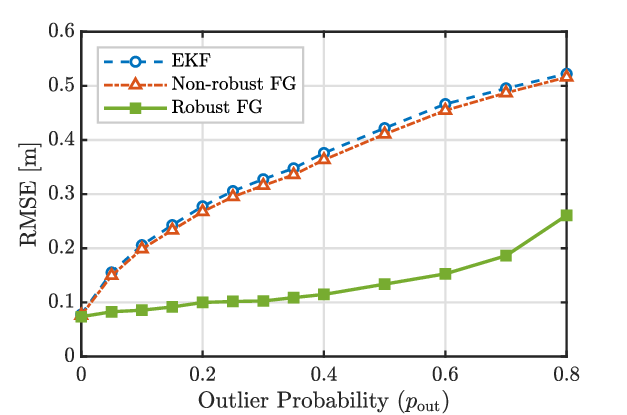}
	\caption{Position RMSE versus outlier probability $p_\text{out}$.}
	\label{fig:rmse}
\end{figure}
\subsection{Evaluation of Robust Localization Under Outliers}

This experiment evaluates the robustness of the proposed factor graph estimator under UWB bearing outliers. The target follows a smooth 3D trajectory. Synthetic UWB measurements are generated with Gaussian noise on range and angles ($\sigma_r=0.05\,\mathrm{m}$, $\sigma_\alpha=\sigma_\beta=3^\circ$). To emulate multipath and NLoS conditions, the nominal bearing noise is replaced, with probability $p_\text{out}$, by a high variance perturbation ($\sigma=25^\circ$), resulting in heavy-tailed angular errors.

Three estimators are compared: (i) an \textbf{Extended Kalman Filter (EKF) baseline} without any outlier handling,
(ii) a \textbf{non--robust factor graph} with standard Gaussian bearing factors, and (iii) the \textbf{proposed robust factor graph} with a Cauchy loss applied to bearing residuals.

For $p_\text{out}=0.2$, Fig.~\ref{fig:sim_traj} shows a representative trajectory. The EKF estimate becomes corrupted by bursts of outliers and drifts from the ground truth. The non--robust factor graph partially mitigates these effects through smoothing but still exhibits noticeable bias. In contrast, the proposed robust factor graph remains close to the true trajectory, demonstrating resilience to corrupted measurements.

Fig.~\ref{fig:rmse} showcases the position Root Mean Square Error (RMSE) over 50 Monte--Carlo runs as a function of $p_\text{out}$. The proposed robust factor graph consistently achieves the lowest RMSE, confirming that combining smoothing with a robust loss is critical for reliable localization under heavy-tailed UWB angular noise.

\begin{figure}[h]
	\centering
	\includegraphics[trim=30 10 45 0, clip,width=0.5\textwidth]{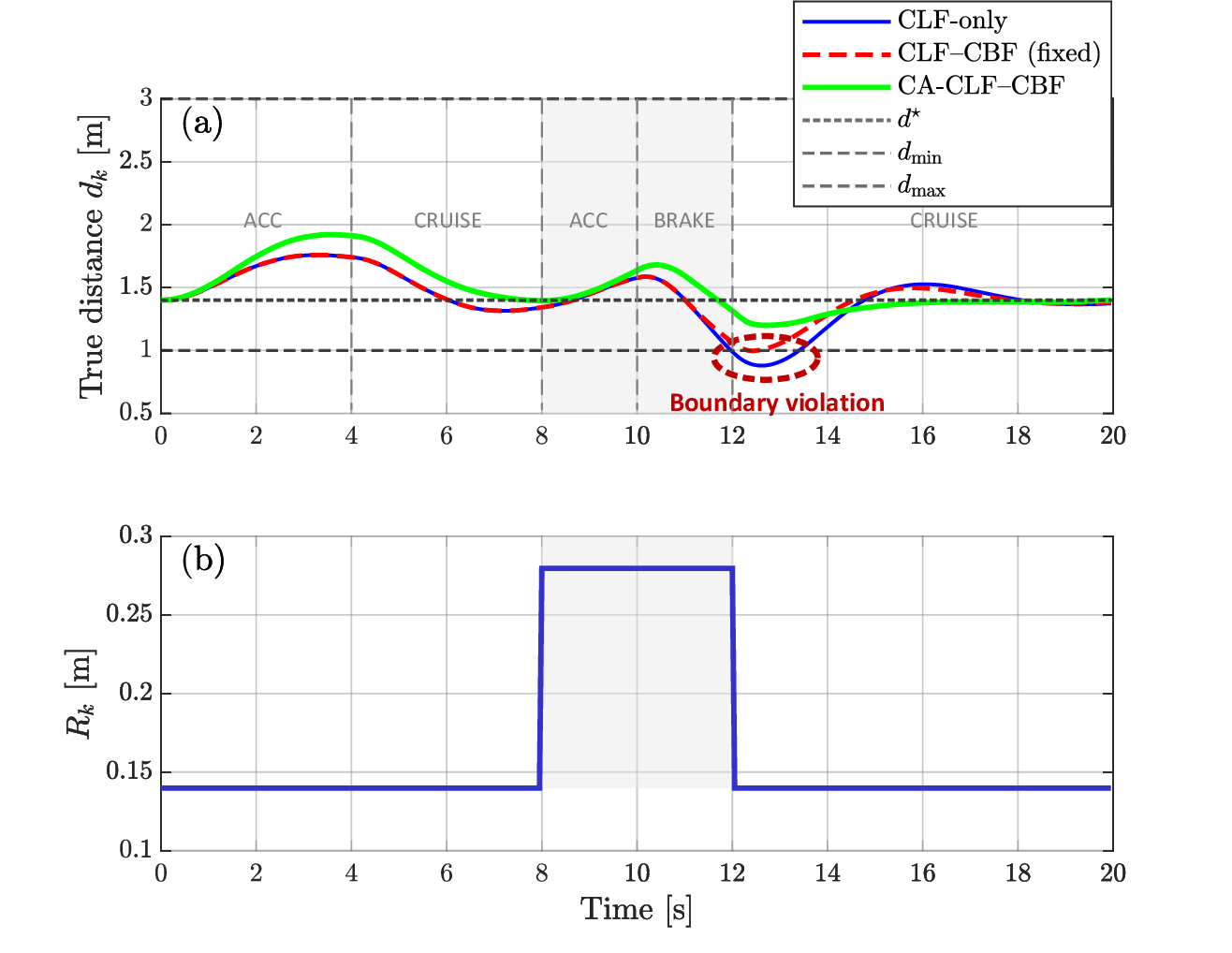}
	\caption{(a) True UAV--target distance under three controllers. 
		(b) Confidence radius $R_k$ derived from the estimator.}
	\label{fig:sim_distance_vs_time}
\end{figure}

\begin{figure*}[t]
	\centering
	\includegraphics[width=0.98\textwidth]{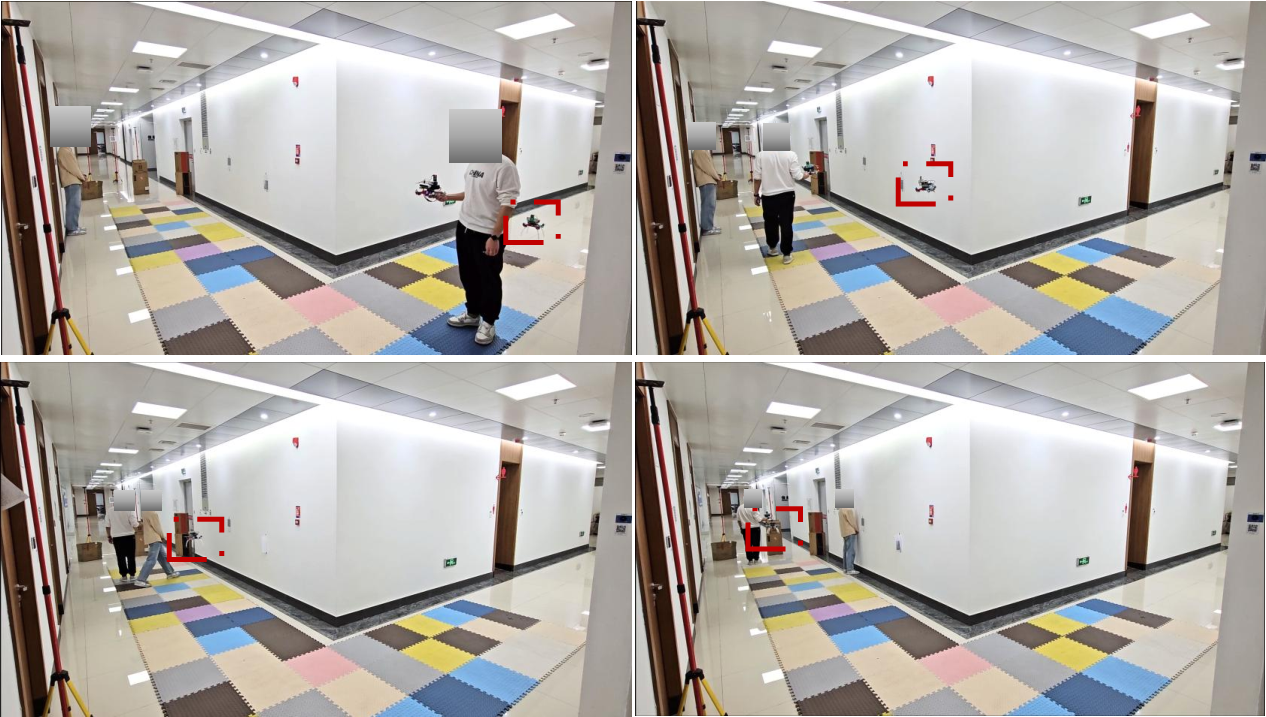}
	\caption{Indoor corridor environment used for real--world validation.}
	\label{fig:Experimental_scenario}
\end{figure*}
\subsection{Evaluation of Covariance-Aware Control Performance}

This experiment evaluates the closed--loop tracking behavior of the proposed covariance-aware controller under varying sensing quality. The target follows a staged motion profile with acceleration, constant--velocity motion, and abrupt deceleration, resulting in non--stationary relative dynamics. To simulate degraded perception, the measurement covariance is increased during a mid--experiment interval, causing a temporary rise in the estimated confidence radius~$R_k$.

Three control strategies are compared: (i) a \textbf{CLF--only} controller without safety enforcement, (ii) a \textbf{fixed CLF--CBF} controller applying constant distance bounds without uncertainty adaptation, and (iii) the proposed \textbf{covariance-aware CLF--CBF} (CA--CLF--CBF) controller.

Fig.~\ref{fig:sim_distance_vs_time} summarizes the results. In subplot~(a), the CLF--only controller exhibits repeated constraint violations, particularly during fast target deceleration. The fixed CLF--CBF controller improves constraint satisfaction, yet still produces occasional boundary breaches when sensing degrades. In contrast, the proposed CA--CLF--CBF controller maintains $d_k$ strictly within the safe bounds for the full duration of the experiment.

Subplot~(b) shows the evolution of $R_k$, which increases only during the degraded--sensing interval. As $R_k$ grows, the controller proactively enlarges the effective safety buffer and increases the temporary standoff distance. This uncertainty--aware behavior avoids aggressive corrections and reduces chattering.

These results demonstrate that incorporating estimation uncertainty into the control barrier constraints significantly improves safety robustness, especially in environments with unreliable sensing.

\begin{figure}[h]
	\centering
	\includegraphics[trim=10 10 10 30, clip,width=0.45\textwidth]{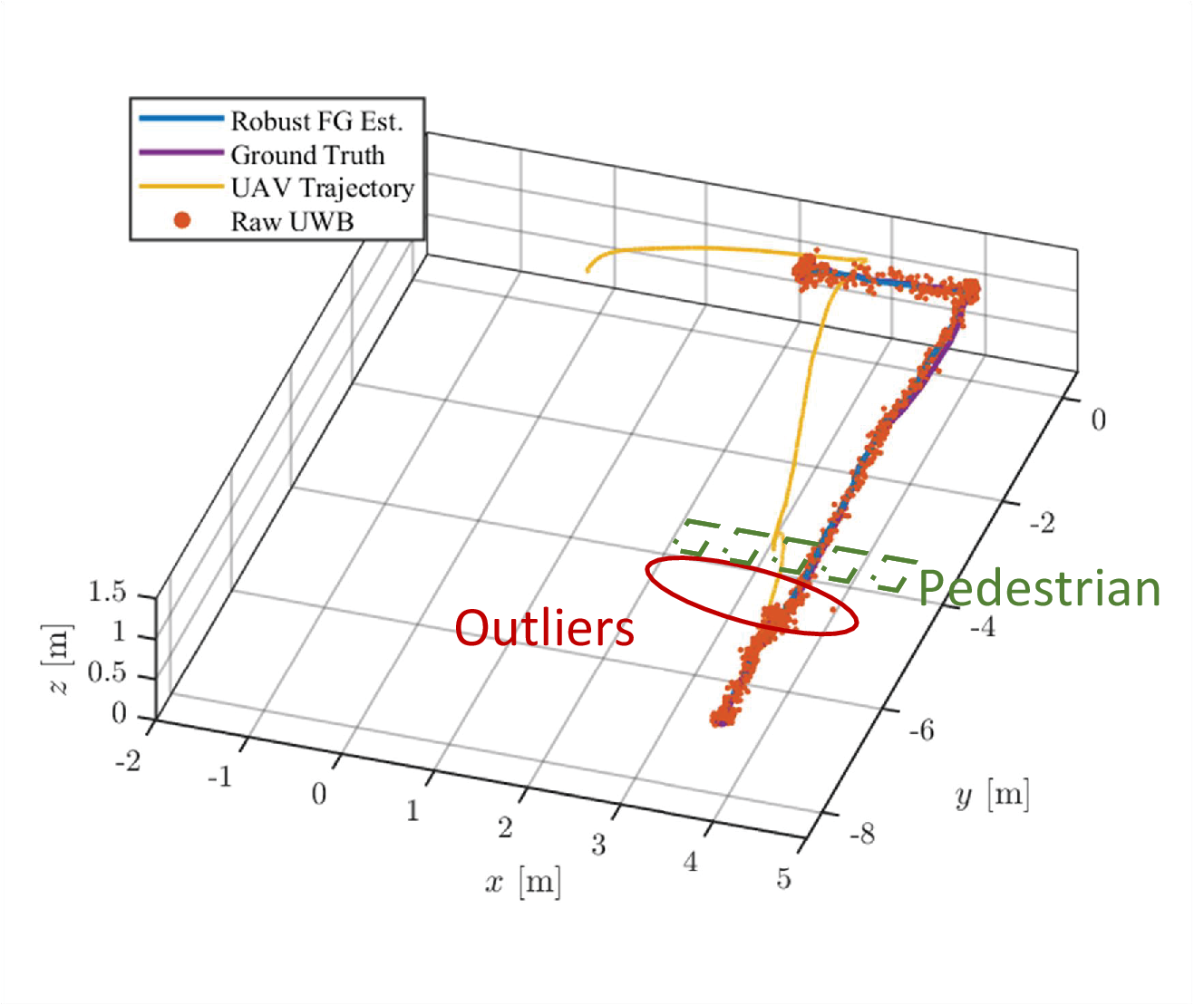}
	\caption{Reconstructed target--following trajectory.}
	\label{fig:trajectory_comparison}
\end{figure}
\section{Experiments}

To evaluate the proposed covariance-aware target localization and CLF--CBF safety--critical tracking framework, we conduct real--world experiments in a narrow indoor corridor characterized by strong multipath conditions and a $90^\circ$ corner. This environment poses significant challenges for UWB--based ranging and bearing. In addition to the geometric constraints imposed by the corridor walls, moving pedestrians are intentionally introduced to create temporary NLoS conditions and abrupt sensing degradation. This setup enables a rigorous evaluation of both the robustness of the estimator and the safety guarantees of the tracking controller. Fig.~\ref{fig:Experimental_scenario} illustrates the experiment scene, where the UAV (highlighted with red brackets) navigates the sharp turn while experiencing dynamic occlusions from pedestrians.

The complete system is deployed on a quadrotor equipped with a single multi-antenna UWB anchor. A NEWRADIOTECH 82885 multi-antenna UWB anchor is rigidly mounted on the UAV and communicates with a corresponding NEWRADIOTECH 82885 UWB tag attached to the moving target. The anchor–tag pair provides raw range and 3D bearing measurements to the onboard robust factor graph estimator. All localization and control modules, including the robust factor graph estimator and the covariance-aware CLF--CBF controller, run in real time entirely onboard an NVIDIA Jetson Orin NX (16 GB) module, forming a fully self-contained estimation–control pipeline that satisfies the computational demands of the proposed framework.

During the experiments, the target moves along a trajectory containing straight segments, a sharp corner turn, and phases of abrupt acceleration and deceleration. The UAV is required to track the target while maintaining a prescribed standoff distance and respecting input and velocity constraints, even under degraded sensing conditions.

Fig.~\ref{fig:trajectory_comparison} shows the reconstructed target--following trajectory. The raw UWB range–bearing measurements contain prominent outliers, especially during periods of pedestrian--induced NLoS. By contrast, the trajectory estimated by the robust factor graph remains smooth, closely aligns with the ground truth, and follows the UAV’s actual flight path. These results validate the estimator’s ability to effectively suppress NLoS-- and multipath--induced measurement errors while accurately recovering the target motion.

\section{Conclusion}
In this letter, we presented an uncertainty--aware single--anchor UWB 3D tracking framework that tightly couples robust relative localization with safety--critical UAV control in confined indoor environments. A robust factor graph formulation fuses UWB range and 3D bearing to deliver target pose and covariance estimates that remain reliable under heavy-tailed noise, multipath, and NLoS--induced outliers. Building on this, a covariance-aware CLF--CBF controller adapts tracking tubes and safety margins according to posterior uncertainty, ensuring safe standoff tracking under actuation, speed, and distance constraints. Numerical simulations and real--world UAV experiments in a narrow indoor corridor confirm that the proposed approach runs fully onboard, maintains safe and stable target following, and systematically adapts to time--varying sensing quality.

\bibliographystyle{IEEEtran}
\bibliography{reference}

\end{document}